\documentclass[%
reprint,
superscriptaddress,
 amsmath,amssymb,
 prx,
]{revtex4-2}
\usepackage{tikz}
\usepackage{graphicx}
\usepackage{amsmath}
\usepackage{gensymb}
\usepackage[margin=1cm]{geometry}
\usepackage{setspace}
\usepackage{xcolor}
\usepackage{cancel}
\usepackage{soul}
\usepackage{wrapfig}
\usepackage{ulem}
\usepackage{units}
\usepackage{lipsum}

\newcommand{\TODO}[1]{\textcolor{black}{#1}}

\newcommand{\reftextit}[1]{}

\renewcommand{\vec}{\mathbf}

\begin{document}

\title{Quantum Dynamics of Attractive and Repulsive Polarons in a Doped MoSe$_2$ Monolayer}

\affiliation{Department of Physics and Center for Complex Quantum Systems, 
    The University of Texas at Austin, Austin, TX, 78712, USA.}
\affiliation{School of Physics and Astronomy and ARC Centre of Excellence in Future Low-Energy Electronics Technologies, Monash University, Victoria 3800, Australia}
\affiliation{Department of Physics, Florida International University, Miami, Florida 33199, USA.}
\affiliation{Microelectronics Research Center, Department of Electrical and Computer Engineering, The University of Texas at Austin, Austin, TX, 78712, USA.}
\affiliation{MONSTR Sense Technologies, LLC, Ann Arbor, Michigan, 48104, USA.}
\affiliation{Research Center for Functional Materials, National Institute for Materials Science, 1-1 Namiki, Tsukuba 305-0044, Japan}
\affiliation{International Center for Materials Nanoarchitectonics, National Institute for Materials Science, 1-1 Namiki, Tsukuba 305-0044, Japan}

\author{Di Huang}
\thanks{These authors contributed equally to this work.}
\affiliation{Department of Physics and Center for Complex Quantum Systems, 
    The University of Texas at Austin, Austin, TX, 78712, USA.}%
\author{Kevin Sampson}
\thanks{These authors contributed equally to this work.}
\affiliation{Department of Physics and Center for Complex Quantum Systems, 
    The University of Texas at Austin, Austin, TX, 78712, USA.}
\author{Yue Ni}
\thanks{These authors contributed equally to this work.}
\affiliation{Department of Physics and Center for Complex Quantum Systems, 
    The University of Texas at Austin, Austin, TX, 78712, USA.}
\author{Zhida Liu}
\affiliation{Department of Physics and Center for Complex Quantum Systems, 
    The University of Texas at Austin, Austin, TX, 78712, USA.}
\author{Danfu Liang}
\affiliation{Department of Physics, Florida International University, Miami, Florida 33199, USA.}
\author{Kenji Watanabe}
\affiliation{Research Center for Functional Materials, National Institute for Materials Science, 1-1 Namiki, Tsukuba 305-0044, Japan}
\author{Takashi Taniguchi}
\affiliation{International Center for Materials Nanoarchitectonics, National Institute for Materials Science, 1-1 Namiki, Tsukuba 305-0044, Japan}
\author{Hebin Li}
\affiliation{Department of Physics, Florida International University, Miami, Florida 33199, USA.}
\author{Eric Martin}
\affiliation{MONSTR Sense Technologies, LLC, Ann Arbor, Michigan, 48104, USA.}
\author{Jesper Levinsen}
\affiliation{School of Physics and Astronomy and ARC Centre of Excellence in Future Low-Energy Electronics Technologies, Monash University, Victoria 3800, Australia}
\author{Meera M. Parish}
\affiliation{School of Physics and Astronomy and ARC Centre of Excellence in Future Low-Energy Electronics Technologies, Monash University, Victoria 3800, Australia}
\author{Emanuel Tutuc}
\affiliation{Microelectronics Research Center, Department of Electrical and Computer Engineering, The University of Texas at Austin, Austin, TX, 78712, USA.}
\author{Dmitry K. Efimkin}
\affiliation{School of Physics and Astronomy and ARC Centre of Excellence in Future Low-Energy Electronics Technologies, Monash University, Victoria 3800, Australia}
\author{Xiaoqin Li}%
\email{elaineli@physics.utexas.edu}
\affiliation{Department of Physics and Center for Complex Quantum Systems, 
    The University of Texas at Austin, Austin, TX, 78712, USA.}
\date{\today}

\begin{abstract}
When mobile impurities are introduced and coupled to a Fermi sea, new quasiparticles known as  Fermi polarons are formed. There are two interesting, yet drastically different regimes of the Fermi polaron problem: (I) the attractive polaron (AP) branch, connected to pairing phenomena spanning the crossover from BCS superfluidity to the Bose-Einstein condensation of molecules; and (II) the repulsive branch (RP), which underlies the physics responsible for Stoner's itinerant ferromagnetism. Here, we study Fermi polarons in two dimensional systems, where many questions and debates regarding their nature persist. The model system we investigate is a doped MoSe$_2$ monolayer. We find the observed AP-RP energy splitting and the quantum dynamics of attractive polarons agree with the predictions of polaron theory. As the doping density increases, the quantum dephasing of the attractive polarons remains constant, indicative of stable quasiparticles, while the repulsive polaron dephasing rate increases nearly quadratically. 
The dynamics of Fermi polarons are of critical importance for understanding the pairing and magnetic instabilities that lead to the formation of rich quantum phases found in a wide range of physical systems including nuclei, cold atomic gases, and solids. 
\end{abstract}

\maketitle

\section{Introduction}
\begin{figure*}
    \centering
    \includegraphics{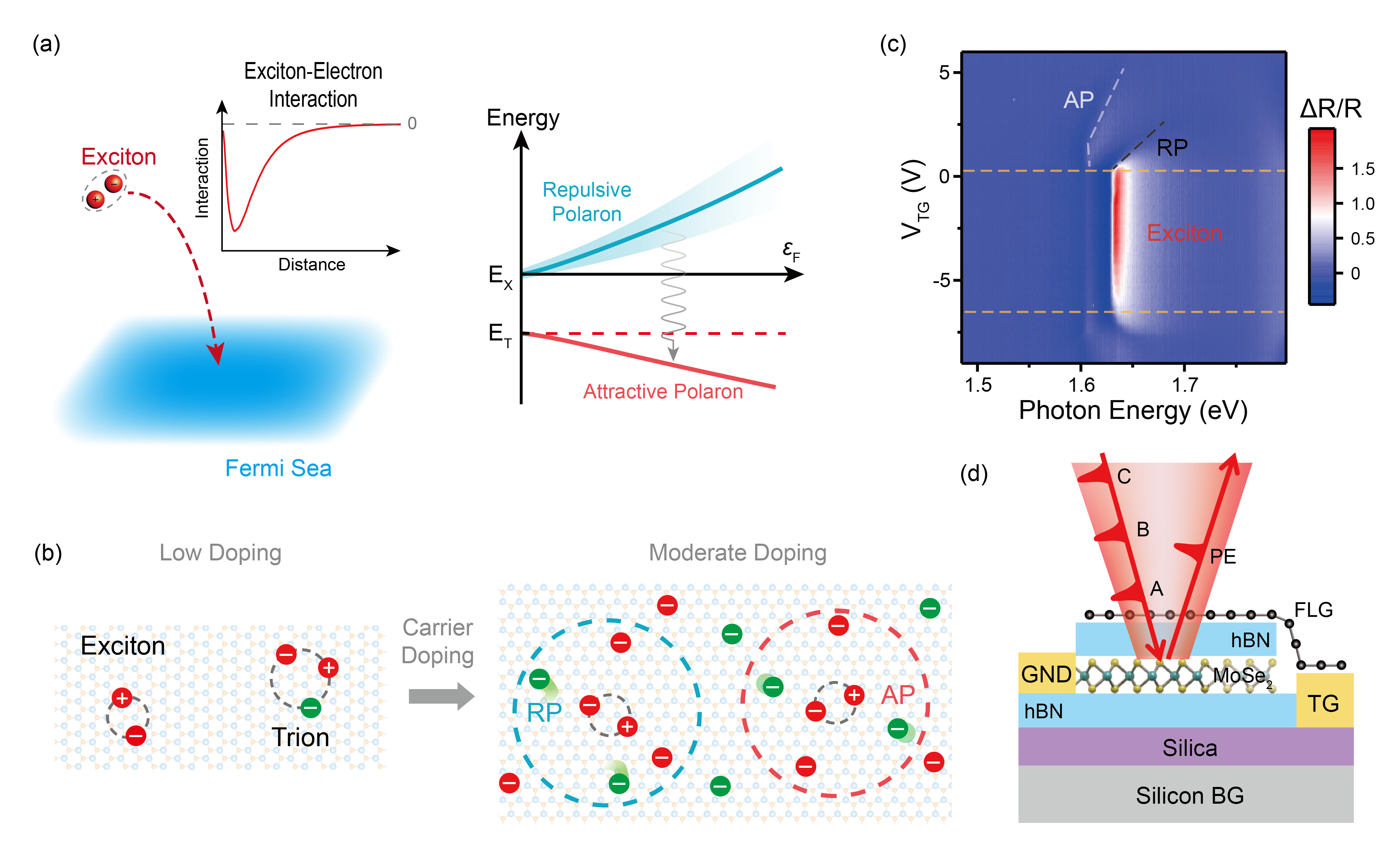}
    \caption{Illustration of the MoSe$_2$ device and physical concept. 
    (a) Schematic of Fermi polarons. When excitons are attractively coupled to a Fermi sea of electrons, the spectrum of the whole system splits into two branches: attractive polarons (APs) and repulsive polarons (RPs). In the limit of vanishing Fermi energy $\varepsilon_{\mathrm{F}}$, the APs and RPs recover the energies of trions ($\mathrm{E_T}$) and excitons ($\mathrm{E_X}$), respectively. 
    (b)  Evolution from an exciton-trion picture to a Fermi polaron picture as the doping density increases in a monolayer. Spheres in red refer to electrons (-) and holes (+) in one valley, while spheres in green color indicate the opposite valley.
    The displacement of electrons around the exciton in the \TODO{moderate-doping} regime represents the polaronic cloud of density fluctuations (particle-hole excitations).
    (c) Reflectance spectra from a doped MoSe$_2$ monolayer at different top gate voltages ($\mathrm{V_{TG}}$). The dashed lines are guidelines for the evolution of APs (bright) and RPs (dark). (d) Schematics of a collinear 2DCES experiment performed on a monolayer MoSe$_2$ device. A train of three pulses (A/B/C) is focused onto the monolayer MoSe$_2$, and the photon echo (PE) signal is collected via the same microscope objective. The pulses are incident perpendicular to the sample but are illustrated here at an angle for clarity.}
    \label{fig:fig1}
\end{figure*}

Fermi polaron quasiparticles are mobile impurities (e.g., excitons) that are coherently dressed by density fluctuations (particle-hole excitations) of a surrounding Fermi sea~\cite{2014_Massignan_RepProgPhys_Review,FermiGasesReview}. As illustrated in Fig.~\ref{fig:fig1}a, the attractive interaction between the exciton and Fermi sea leads to an energetically favorable state --- the attractive polaron --- as well as a higher energy repulsive polaron, a metastable state that eventually 
decays into attractive polarons. 
The behavior of Fermi polarons has been extensively studied in the context of ultracold atomic gases, where experiments have probed the formation of polarons~\cite{Cetina2016} as well as their quasiparticle properties, such as the polaron energy\cite{Schirotzek2009}, lifetime~\cite{2012_C.Kohstall_Nature_FermiPolColdAtom}, and the effective mass~\cite{2009_Salomon_PRL_3DFermiPolaron}. 
However, cold-atom experiments have focused on three dimensional systems. Intriguing and open questions remain, e.g., how the Fermi polarons evolve as the dimensionality is lowered and low-energy density fluctuations of the Fermi sea are enhanced. In one dimensional systems, the impurity problem becomes exactly solvable~\cite{McGuire1966}, and polaronic quasiparticles are destroyed by quantum fluctuations. Thus, the intermediate case of two dimensional (2D) systems represents a particularly interesting yet unresolved case, where quantum fluctuations are 
enhanced 
but quasiparticles can still exist.

Atomically thin transition metal dichalcogenides (TMDCs) represent exemplary 2D systems to study the rich properties of  Fermi polarons, where an exciton is dressed by Fermi seas with a unique valley degree of freedom~\cite{2016_A.Imamoglu_NatPhys_MoSe2,2017_Dmitry_PRB_theory}. A variety of exciton resonances (neutral or charged states, biexcitons, spatial and momentum-space indirect excitons) have been identified and investigated in TMDC monolayers~\cite{2018_M.Ermin_TMD_Exciton_Review, 2020_He_NatComm_ExcitonComplexes}.  The trion, a three-body bound state consisting of an exciton bound to an extra electron or hole, has been used widely in the literature to describe an optical resonance appearing at an energy $\sim$ 20-30 meV below the neutral excitons~\cite{2013_KF.Mak_Nat.Mat._Trion}. The trion picture is difficult to distinguish from 
the polaron picture at low doping densities, where the Fermi energy is smaller than the intrinsic exciton linewidth~\cite{2020_Glazov_JCP_EqualTrionPolaron,TrionDoubtSuris,TrionDoubtCombescot,TrionDoubtWouters2}.
However, as doping increases, a continuous energy shift suggests that the three-body bound state picture is insufficient~\cite{2022_S.Crooker_NanoLett_WSe2ManyBody, 2020_A.Chernikov_PRL_WSe2Trion, 2022_S.Crooker_hexcitons&oxciton}.  Attractive and repulsive polarons (APs and RPs), as illustrated in Fig.~\ref{fig:fig1}b, have been proposed to explain the evolution of reflectivity spectra of doped TMDC monolayers and bilayers with doping~\cite{2016_A.Imamoglu_NatPhys_MoSe2, 2019_A.Imamoglu_PRL_PolaronMagField}. This polaron picture is anticipated to be relevant up to moderate doping, where the interelectron distance is sufficiently large compared to the 
radii of photoexcited excitons such that their formation is not disturbed. A critical question remains: does polaron theory predict different properties of APs and RPs?  Quantum decoherence associated with APs and RPs, for example, has never been investigated experimentally. Thus, a comparison between experiments and theory has not been possible so far.
  
Here, we study the emergence and evolution of APs and RPs in a MoSe$_2$ monolayer as the electron doping density increases. Using two-dimensional coherent electronic spectroscopy (2DCES), we follow the changes in resonant energy, oscillator strength, and quantum decoherence of the AP and RP branches. The observed red-shift of the AP resonance suggests that these are new and energetically stable quasiparticles. The red-shift increases until a critical doping density is reached and a more pronounced blue shift occurs. Intriguingly, the quantum decoherence rate of APs (measured via the homogeneous linewidth) remains nearly constant up to a critical density, which suggests that the excitons are coherently dressed by the Fermi sea. The quantum decoherence rate of RPs, however, increases monotonically and rapidly with  doping density. Our study highlights that optically excited TMDCs represent a novel playground to study the rich phenomenology of strongly imbalanced quantum mixtures. 

\section{Results}
We control the doping density in a MoSe$_2$ monolayer encapsulated between hBN layers using a device with a few-layer graphite top gate (TG) and a silicon back gate (BG) (details in Supplemental Material). All optical measurements are taken at \unit[16]{K} unless stated otherwise. We firstly measure reflectance spectra at different doping levels, as shown in Fig.~\ref{fig:fig1}c. When the MoSe$_2$ monolayer is charge neutral or in the regime of dilute doping density (between the two orange dashed lines), two resonances at \unit[1635]{meV} and \unit[1605]{meV} are attributed to excitons and trions, respectively. We focus on the electron doping regimes with  positive TG voltages. The exciton resonance evolves into the repulsive polaron (RP) branch, which exhibits a blue-shift at $\mathrm{V_{TG}}> \unit[0.5]{V}$. The lower energy resonance exhibits a small red-shift for $\mathrm{V_{TG}} = \unit[0.5\sim1.5]{V}$ before a pronounced blue shift starts to occur at $\mathrm{V_{TG}}>\unit[1.5]{V}$. 

\begin{figure*}
    \centering
    \includegraphics{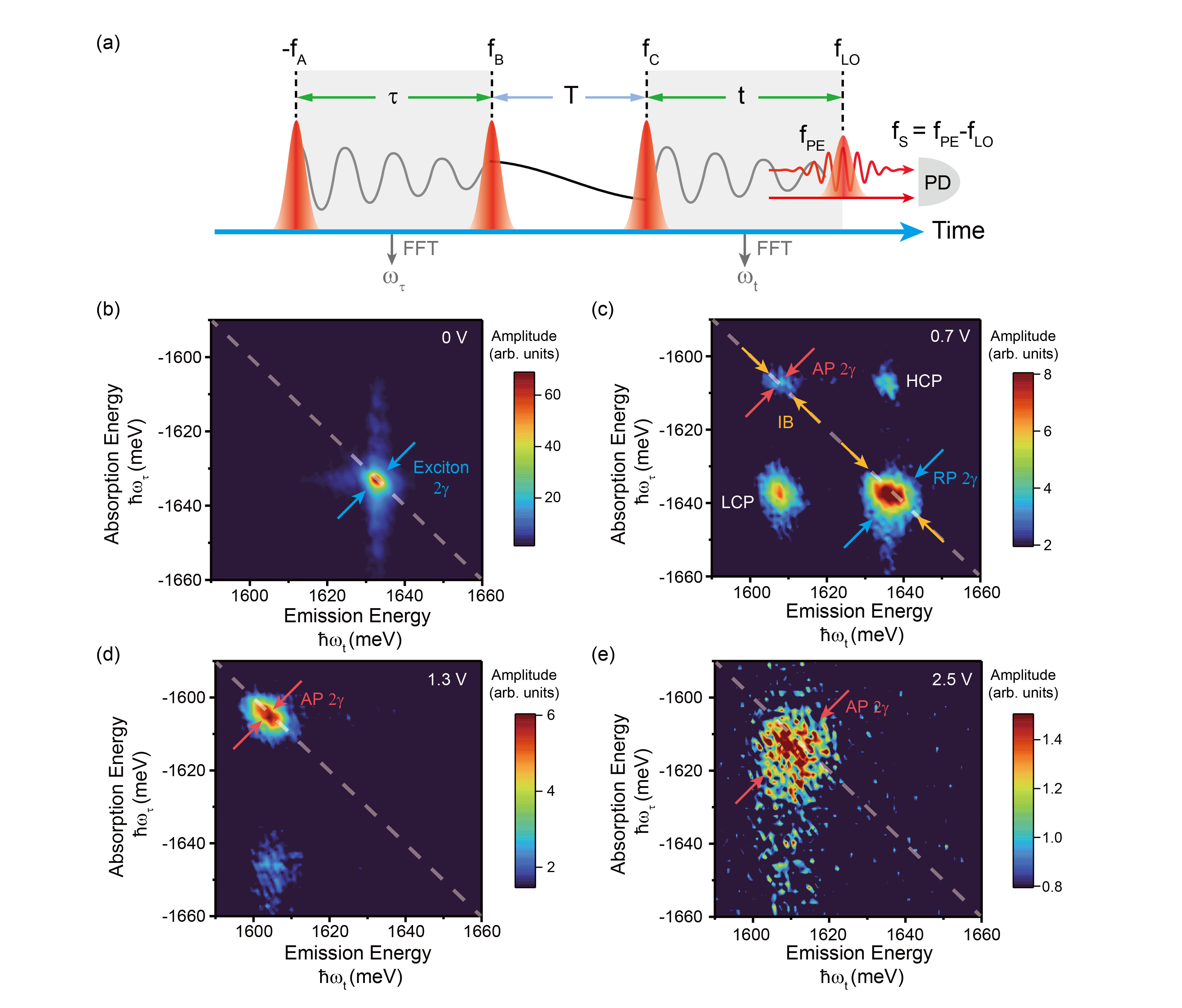}
    \caption{One-quantum rephasing amplitude spectra of the MoSe$_2$ monolayer at different gate voltages. (a) Schematic of the one-quantum rephasing pulse sequence. Details are further discussed in the Supplemental Material. (b-e) 2D spectra taken at different top gate voltages at \unit[0]{V}, \unit[0.7]{V}, \unit[1.3]{V} , and \unit[2.5]{V} respectively. These gate voltages correspond to estimated electron doping densities of $n_{e^-}=0, 6.6\times10^{11}, 2.6\times10^{12}$, and $\unit[6.6\times10^{12}] {e^-/cm^2}$. IB: inhomogeneous broadening; HCP/LCP: higher/lower off-diagonal cross peak. Homogeneous (inhomogeneous) linewidth of exciton, AP, and RP could be extracted by fitting the cross-diagonal (diagonal) slices of the spectra, as illustrated by the arrows in blue (exciton/RP) and red (AP) color. }
    \label{2DCS}
\end{figure*}

In the following, we analyze the spectral shift, oscillator strength changes, and quantum dephasing of these resonances using 2DCES. As illustrated in Fig.~\ref{fig:fig1}d, three collinearly polarized ultrafast pulses derived from the same laser are focused onto the sample by a microscope objective. The third-order optical nonlinear response leads to a coherent photon echo signal, which is collected by the same objective, isolated and detected using a frequency modulation scheme after the heterodyne detection with a reference pulse (details in Supplemental Material). The laser spectrum covers both the AP and RP resonances identified in the reflectance spectrum, and resonantly excites them. 

We perform one-quantum rephasing measurements in which the time delays between the first two excitation pulses ($\tau$) and that between the third pulse and a reference pulse ($t$) are scanned, as illustrated in Fig.~\ref{2DCS}a. These two time delays ($\tau$ and $t$) are then Fourier transformed to yield the absorption energy ($\hbar\omega_{\tau}$) and emission energy ($\hbar\omega_{t}$), respectively. The waiting time $T$ between the second and third pulse is kept at 100 fs to avoid artifacts that occur during the temporal overlap between the excitation pulses. The amplitude of the photon echo signal as a correlation between absorption and emission frequencies is shown in Figs.~\ref{2DCS}b-e. The elongation along the diagonal direction indicated by the dashed line is determined by the inhomogeneous linewidth while the full width half maximum (FWHM) along the cross diagonal direction indicated by two arrows reveals the homogeneous linewidth $2\gamma$~\cite{2010_Mark_OE_2DCSLineShape}, where $\gamma$ is the decoherence rate and inversely proportional to the quantum dephasing time $1/T_2=\gamma/\hbar$.

\begin{figure*}
    \centering
    \includegraphics{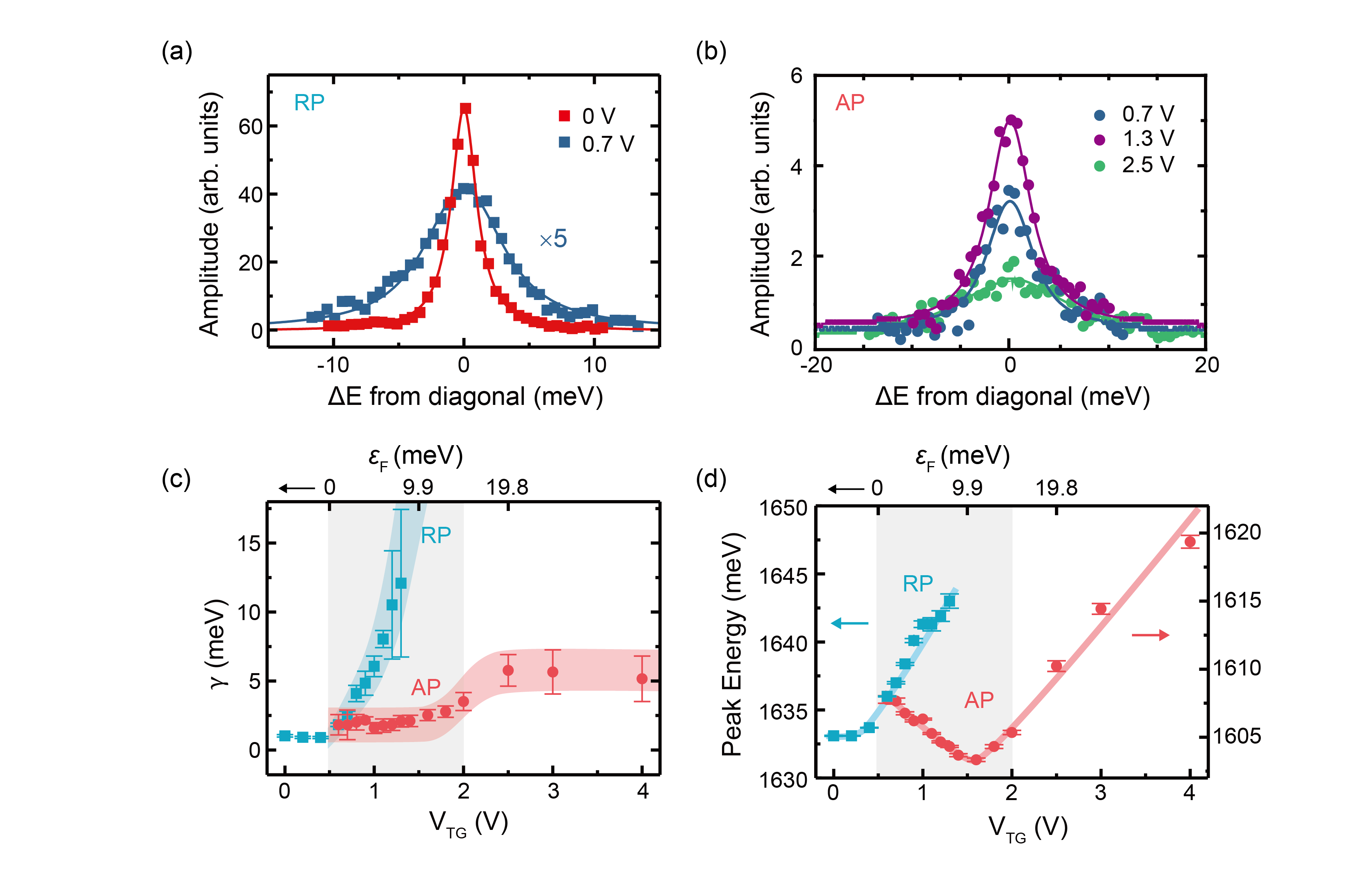}
    \caption{Homogeneous linewidths and resonant energies of APs and RPs extracted from 2D spectra. Cross-diagonal line cuts of 2D spectra used to extract homogeneous linewidths for (a) RPs and (b) APs at a few selected top gate voltages. The data points are fitted by a modified Voigt function as discussed in the Supplemental Material. (c) Homogeneous linewidths and  (d) Resonant energy of RPs (blue) and APs (red) as a function of top gate voltage (bottom X-axis) and corresponding Fermi levels (top X-axis). The shaded grey area indicates the doping density range over which the polaron theory applies.}
    \label{analysis}
\end{figure*}

We further investigate the one-quantum spectra of monolayer MoSe$_2$ at several doping levels in Figs.~\ref{2DCS}b-e. In the charge neutral regime up to $\mathrm{V_{TG}}=\unit[0.5]{V}$, the spectrum is dominated by the exciton resonance at $\sim\unit[1633]{meV}$~\cite{2015_G.Moody_NatComm_2DCSMoSe2, 2020_E.Martin_PRApp_2DCSMoSe2}. At the carrier density of $n=\unit [6.6\times10^{11}] {e^-/cm^2}$ (or $\mathrm{V_{TG}}=\unit[0.7]{V}$), both AP and RP are observed as two diagonal peaks appearing at $\sim\unit[1605]{meV}$ and  $\sim\unit[1637]{meV}$, respectively. Moreover, two off-diagonal cross peaks emerge, which are labeled lower/higher cross peak (i.e., LCP and HCP). These cross-peaks result from electronic coupling between the APs and RPs as previously studied in the nominally doped MoSe$_2$ monolayers~\cite{2016_H.Kai_NanoLett_CP-2DCS}. The unbalanced LCP and HCP intensities originate from both coherent quantum beats and decays from RP to AP. Upon further increasing the doping density to $n=\unit [2.6\times10^{12}] {e^-/cm^2}$ ($\mathrm{V_{TG}}=\unit[1.3]{V}$), the spectral weight has nearly completely shifted to APs. A weak LCP peak remains in the spectrum, suggesting that the metastable RPs still absorb light but quickly decay to APs. At the highest doping density $n=\unit [6.63\times10^{12}] {e^-/cm^2}$ ($\mathrm{V_{TG}}=\unit[2.5]{V}$) where a nonlinear signal is still detectable, the AP resonance becomes much broader and shifts to higher energy.

We carefully analyze the energy shifts and quantum decoherence rates of APs and RPs extracted from 2D spectra as shown in Fig.~\ref{analysis}. The slices along the cross-diagonal direction for the RPs and APs peak at a few selected TG voltages are displayed in Fig.~\ref{analysis}a and Fig.~\ref{analysis}b, respectively. The homogeneous linewidth of the RP rapidly increases with electron doping density, ($\gamma=\unit[1.03\pm0.09]{meV}$ and $\unit[2.50\pm0.30]{meV}$ with $\mathrm{V_{TG}}=\unit[0]{V}$ and $\unit[0.7]{V}$, respectively) corresponding to higher decoherence rate as summarized in Fig.~\ref{analysis}c (blue curve).  By contrast, the decoherence rate of APs remains largely constant until the doping density exceeds a critical density at  $\mathrm{V_{TG}}=\unit[1.5]{V}$  as shown in Fig.~\ref{analysis}c (red curve), ($\gamma=\unit[1.87\pm1.06]{meV}$, $\unit[2.08\pm0.41]{meV}$ and $\unit[5.77\pm1.14]{meV}$ with $\mathrm{V_{TG}}=\unit[0.7]{V}$, $\unit[1.3]{V}$ and $\unit[2.5]{V}$, respectively.) Fig.~\ref{analysis}d displays the central energies of the APs (red points) and RPs (blue points) which are extracted from fitting to the inhomogeneous slices along the diagonal direction of the 2D spectra (details in Supplemental Material Fig.~S5 and S6).
In the range of gate voltages that correspond to a modest doping density, a continuous red-shift of the AP resonance is observed, suggesting the coherent dressing of the quasiparticle can lower its energy. The red-shift of the APs against doping level is often overlooked or left unexplained in previous studies~\cite{2021_CHLui_NatComm_MoSe2&WSe2}. By contrast, the energy of the RP exhibits a blue shift over the whole doping range in which this resonance is observable in the 2DCES spectra. Interestingly, the energy of the APs begins to blue shift at a critical density that approximately coincides with where the quantum dephasing rate starts to increase. 

\section{Discussion and Conclusion}

\begin{figure*}
    \centering
    \includegraphics{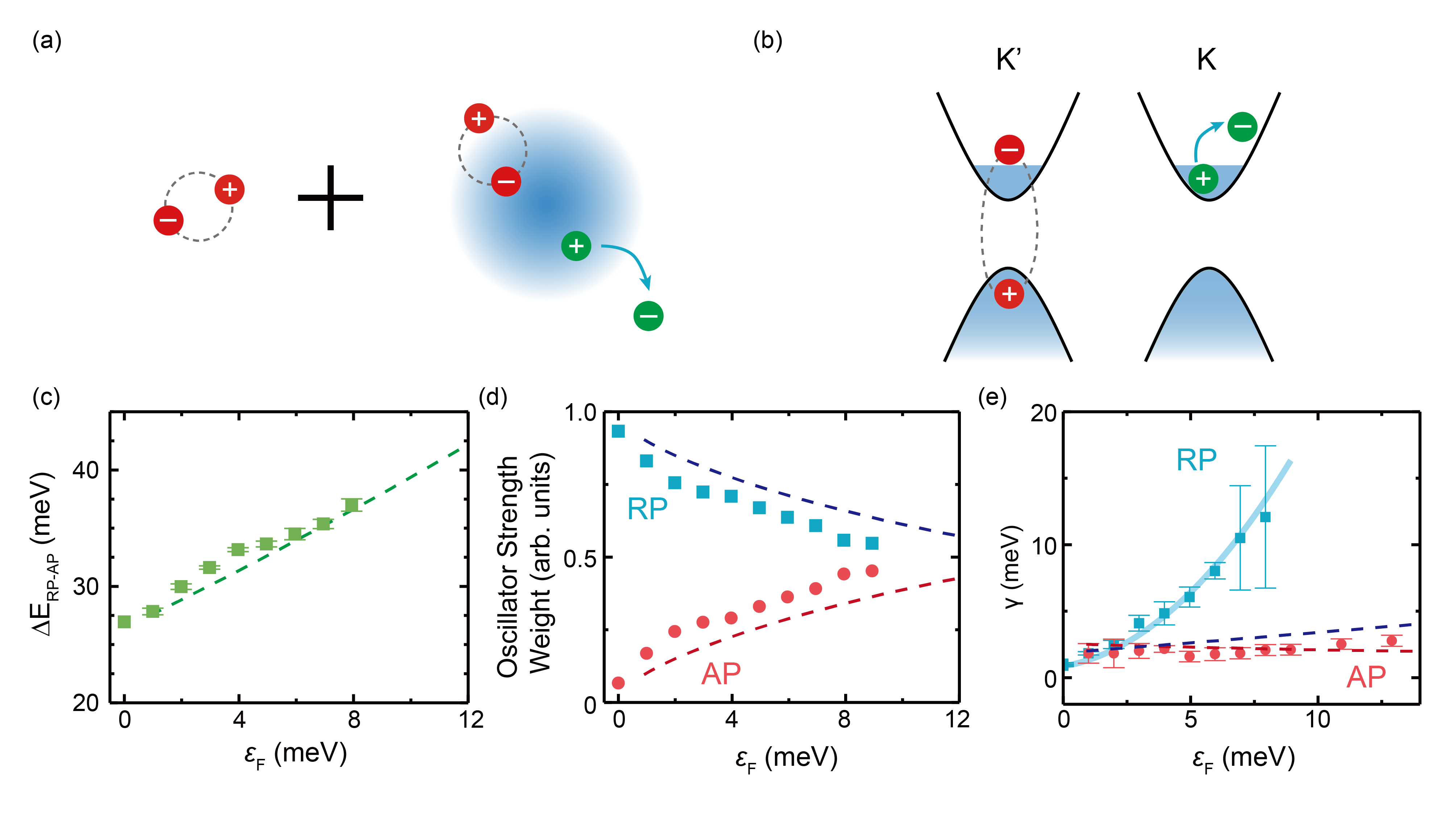}
    \caption{Polaron theory and comparison with experiments. (a) The exciton-polaron is the coherent superposition of a bare exciton and an exciton dressed by a polaron cloud approximated by a single Fermi sea (FS) unbound electron-hole pair in momentum space. (b) Exciton and the FS pair are assumed to reside in different valleys. 
    In panels (a)-(b), red spheres refer to electrons (-) and holes (+) in one valley, while green spheres indicate those in the opposite valley. Comparisons between measurements and calculations for (c) energy splitting between APs and RPs, (d) relative oscillator strengths, and (e) homogeneous linewidths of APs and RPs as a function of Fermi level. In panels (c)-(e), the dashed lines indicate calculations while solid points are extracted from 2D spectra.}
    \label{Theory}
\end{figure*}

Remarkably, most of the key experimental observations (doping dependent energy, oscillator strength, and quantum decoherence rate) can be well captured by a microscopic theory based on the simple Chevy ansatz~\cite{Chevy2006} of Fermi polarons. 
The excess electrons have two main effects: exciton renormalization and polaronic dressing~\cite{2016_A.Imamoglu_NatPhys_MoSe2}. The former modifies the resonance frequency and oscillator strength of excitons, but does not lead to polaron formation or two separate branches. These excitons interact with excess electrons and the polaronic dressing splits them into AP and RP branches. The polaronic physics is solely responsible for the relative behavior of the two branches (i.e., the relative frequency, broadening and the oscillator strength), which is the focus of the present paper.  As illustrated in Fig.~\ref{Theory}a, an exciton-polaron ($P^\dagger_{\vec{q}}$) represents the coherent superposition of the undressed exciton ($X_{\vec{q}}^\dagger$ with weight $\phi_{\vec{q}}$) and the polaron cloud, which is approximated as a single electron-hole excitation of the opposite valley to that in which the exciton resides ($f_{\vec{k}}^\dagger f_{\vec{k}'}$ in the Fermi sea with weight $\chi_{\vec{q kk}'}$ and momenta $k>k_\mathrm{F}$ and $k'<k_\mathrm{F}$): 
\begin{equation}
\label{PF}
P^\dagger_{\vec{q}}=\phi_{\vec{q}} X_{\vec{q}}^\dagger + \sum_{\vec{k}\vec{k}'} \chi_{\vec{q} \vec{k} \vec{k}'} X_{\vec{q}+\vec{k}'-\vec{k}}^\dagger f_{\vec{k}}^\dagger f_{\vec{k}'}.
\end{equation}
As sketched in Fig.~1b and Fig.~\ref{Theory}b, the exciton and the unbound electron-hole excitation responsible for the polaronic dressing reside in different valleys in our microscopic theory, since same-valley correlations are suppressed by exchange effects due to the indistinguishability between the electron bound in the exciton and those in the Fermi sea~\cite{Tiene2022}. 

The comparisons between experimental observations and predictions from the Fermi-polaron theory are summarized in Figs.~\ref{Theory}c-e. 
A hallmark of the Fermi-polaron theory in 2D is a linear increase of the energy splitting between APs and RPs ($\Delta E_{\mathrm{RP-AP}}$) as a function of Fermi energy as $\Delta E_{\mathrm{RP-AP}}=\varepsilon_{\mathrm{T}}+3\varepsilon_{\mathrm{F}}/2$ where $\varepsilon_{\mathrm{T}}$ is the binding energy of the exciton-electron bound state (i.e., the trion) and is treated as a fitting parameter.
The prefactor $3/2$ originates from the inverse reduced exciton-electron mass $2m_e/3$ in TMDCs~\cite{2021_Dmitry_PRB_theory}, and is a signature of a mobile impurity. 
The experimentally extracted energy splitting matches the prediction remarkably well (Fig.~\ref{Theory}d) including the slope of the linear dependence~\footnote{Our calculations follow Ref.~\cite{2017_Dmitry_PRB_theory}. In the recent paper~\cite{2021_Dmitry_PRB_theory}, exciton-electron interactions and the binding energy $\varepsilon_{\mathrm{T}}$ are calculated microscopically and $\varepsilon_{\mathrm{T}}$ reasonably agrees with the trion binding energy evaluated within the genuine three-particle problem. Importantly, the doping dependence of absorption calculated within the two models (with microscopically derived interactions and ones approximated by the contact pseudopotential), are very close to each other.}. The parameters used in the analysis of experimental data are discussed in detail in Supplemental Material, in particular, $\varepsilon_{\mathrm{T}}$ is estimated to be $\sim\unit[26.9]{meV}$ in our fitting. We further analyze the relative oscillator strength transfer between the APs ($f_{\mathrm{AP}}$) and RPs ($f_{\mathrm{RP}}$)  as a function of the Fermi level. The oscillator strength of both APs and RPs is evaluated by integrating the amplitude over the area (labeled as $S_{\mathrm{AP}}$ and $S_{\mathrm{RP}}$) around the resonances in the 2DCES spectra (details in Supplemental Material). An excellent agreement is found between our measurements and the prediction of the polaron theory in Fig.~\ref{Theory}e. \textcolor{black}{Note that the oscillator strength of the AP peak vanishes as the Fermi energy goes to zero, while the energy splitting between branches remains finite and reduces to the trion binding energy. Moreover, for low doping $\varepsilon_{\mathrm{F}} \lesssim 2$ meV, the trion and polaron pictures give equivalent results for the oscillator strength~\cite{2020_Glazov_JCP_EqualTrionPolaron}.}

\textcolor{black}{Another key prediction of the Fermi-polaron theory is} that the AP linewidth is almost doping independent, which agrees well with the experimental observation up to $\varepsilon_{\mathrm{F}}$ $\sim$ \unit[15]{meV} (Fig.~\ref{Theory}d). \textcolor{black}{This behavior cannot be captured within the trion picture, which instead predicts that the AP linewidth increases approximately linearly with $\varepsilon_{\mathrm{F}}$~\cite{Zhang2014,2020_Glazov_JCP_EqualTrionPolaron}}. However, the \textcolor{black}{polaron} theory substantially underestimates the observed RP broadening as doping density increases. This observation is in sharp contrast to the situation in ultracold atomic gases, where the Chevy ansatz accurately describes the RP linewidth~\cite{Adlong2020}. The observed RP linewidth is very well fitted by adding an extra quadratic term to the linear dependence on $\varepsilon_{\mathrm{F}}$ predicted by the polaron theory (blue solid line in Fig.~\ref{Theory}e). This discrepancy suggests the existence of a qualitatively new contribution to RP decay. 
The quadratic dependence of the additional broadening hints at the involvement of electron-electron interactions. In particular, the additional RP decay can originate from nonradiative transitions from the RP to the AP. Although we cannot rule out phonons and charge fluctuations as possible contributing factors to dephasing, they are anticipated to affect both polaronic branches in similar ways, making them unlikely reasons for different doping-dependent dephasing. The decay due to  electron-electron interactions involves an extra particle-hole pair that carries away the energy difference between the branches, leading to a six-particle final state that is not included in the Chevy ansatz. This decay process is enhanced when the Fermi-sea pair is in the same valley as the exciton since this avoids Pauli blocking effects with the dressing cloud.    


We estimate the non-radiative decay from RPs to APs using Fermi's golden rule:
\begin{equation}
\begin{split}
\label{Gamma}
\gamma_{\mathrm{ee}}= \pi\sum_{\vec{q} \vec{k}'}&|\langle f |\hat{H}_{e-e}|i\rangle|^2 \\ & \times \delta(
\Delta E_{{\rm RP}-{\rm AP}}-\epsilon^\mathrm{X}_{\vec{q}}-\epsilon^\mathrm{e}_{\vec{k}'-\vec{q}} + \epsilon^\mathrm{e}_{\vec{k}'}).
\end{split}
\end{equation}
Here, $|i\rangle=P^\dagger_{\vec{0},\mathrm{R}}|g\rangle$ is the initial RP state while the final state $|f\rangle=P^\dagger_{\vec{q},\mathrm{A}}\tilde{f}^\dagger_{\vec{k}'-\vec{q}} \tilde{f}_{\vec{k}'}  |g\rangle$ 
includes an electron-hole excitation ($|\vec{k}'-\vec{q}|>k_\mathrm{F}$ and $k<k_\mathrm{F}$) that is distinguishable from that participating in the polaron state described by Eq.~(\ref{PF}). \textcolor{black}{We neglect short-range exciton-exciton and exciton-electron interactions and keep only the dominant long-range electron-electron Coulomb interactions in $\hat{H}_{e-e}$}. \textcolor{black}{The energy conservation is determined by the energy-difference $\Delta E_{{\rm RP}-{\rm AP}}$ between the zero-momentum RP and AP states, the dispersion of the AP in the final state which can be approximated by the bare exciton dispersion $\epsilon^\mathrm{X}_{\vec{k}}$, and the energy associated with an electron-hole excitation in terms of the electron dispersion $\epsilon^\mathrm{e}_{\vec{k}}$.} 
As presented in the Supplemental Material, the additional broadening can be written as 
\begin{equation}
\label{Gamma2}
\gamma_{\mathrm{ee}}\approx\frac{3\pi M^2}{4} \frac{\varepsilon_{\mathrm{F}}^2\varepsilon_{\mathrm{X}}}{\varepsilon_\mathrm{T}^2}.
\end{equation}
Here $\varepsilon_{\mathrm{X}}$ is the binding energy for excitons. The dimensionless constant $M$ describes the overlap  between polaronic clouds in the initial and final states (details in Supplemental Material). $M$ is largely independent of doping density and has been used as a fitting parameter. Most importantly, this extra contribution is proportional to $\varepsilon_{\mathrm{F}}^2$. When it is combined with the linear doping dependence from the Chevy ansatz,  the experimental data can be fitted well. 

Examples of accurate theory for strongly interacting many-body systems are rare. In this work, we demonstrate how the Fermi polaron theory can be used to describe the quantum dynamics of an imbalanced quantum mixture of excitons and an electron gas in a MoSe$_2$ monolayer. While Fermi polarons in the cold atom systems and 2D semiconductors share similar properties, e.g., the linear RP-AP energy splitting and stable quantum dynamics of APs as a function of Fermi level~\cite{FermiGasesReview}, there are also important differences. The quantum dephasing rate of RPs increases quadratically with the doping density in TMDC monolayers in contrast to the linear dependence found in cold atom systems~\cite{Oppong2019}. The additional broadening is intricately connected with the rich and complicated interplay of two valleys that is present not only in MoSe$_2$, but in all TMDC monolayer semiconductors. Our study demonstrates another fruitful playground to study the Fermi polaron problem complementary to cold atoms where this problem has been investigated intensely in recent years~\cite{2012_C.Kohstall_Nature_FermiPolColdAtom, 2012_M.Koschorreck_Nature_FermiPolColdAtom,Oppong2019,2017_Scazza_PRL_RPColdAtom}. The understanding of Fermi polarons in monolayers also provides a foundation for exploring the coupling between excitons and other correlated electronic ground states such as Wigner crystals, Mott insulating states, and superconductivity-like states in doped TMDC twisted bilayers~\cite{2021_T.Smolenski_Nature_PolaronWignerCrystal}.  


\begin{acknowledgments}
The spectroscopic experiments performed by D. H. and K. S. at UT-Austin were primarily supported by the Department of Energy, Basic Energy Science program via grant DE-SC0019398 and N. Y. is partially supported by NSF DMR-1808042. The work was partly done at the Texas Nanofabrication Facility supported by NSF grant NNCI-2025227. K. S. acknowledges a fellowship via NSF DMR-1747426 and partial support from NSF MRSEC program DMR-1720595, which also supports the facility for preparing the sample. X.L gratefully acknowledges sample preparation support by the Welch Foundation via grant F-1662. H.L. acknowledges support by NSF via grant DMR-2122078. J.L., D.K.E., and M.M.P. acknowledge support from the Australian Research Council (ARC) Centre of Excellence in Future Low-Energy Electronics Technologies (CE170100039). J.L. and M.M.P. are also supported through the ARC Future Fellowships FT160100244 and FT200100619, respectively, and J.L. furthermore acknowledges support from the ARC Discovery Project DP210101652. E.T. acknowledges support from the Army Research Office Grant No. W911NF-17-1-0312, and the NSF MRSEC program DMR-1720595. K.W. and T.T. acknowledge support from JSPS KAKENHI (Grant Numbers 19H05790, 20H00354 and 21H05233).
\end{acknowledgments}

%


\begin{thebibliography}{35}%
	\makeatletter
	\providecommand \@ifxundefined [1]{%
		\@ifx{#1\undefined}
	}%
	\providecommand \@ifnum [1]{%
		\ifnum #1\expandafter \@firstoftwo
		\else \expandafter \@secondoftwo
		\fi
	}%
	\providecommand \@ifx [1]{%
		\ifx #1\expandafter \@firstoftwo
		\else \expandafter \@secondoftwo
		\fi
	}%
	\providecommand \natexlab [1]{#1}%
	\providecommand \enquote  [1]{``#1''}%
	\providecommand \bibnamefont  [1]{#1}%
	\providecommand \bibfnamefont [1]{#1}%
	\providecommand \citenamefont [1]{#1}%
	\providecommand \href@noop [0]{\@secondoftwo}%
	\providecommand \href [0]{\begingroup \@sanitize@url \@href}%
	\providecommand \@href[1]{\@@startlink{#1}\@@href}%
	\providecommand \@@href[1]{\endgroup#1\@@endlink}%
	\providecommand \@sanitize@url [0]{\catcode `\\12\catcode `\$12\catcode
		`\&12\catcode `\#12\catcode `\^12\catcode `\_12\catcode `\%12\relax}%
	\providecommand \@@startlink[1]{}%
	\providecommand \@@endlink[0]{}%
	\providecommand \url  [0]{\begingroup\@sanitize@url \@url }%
	\providecommand \@url [1]{\endgroup\@href {#1}{\urlprefix }}%
	\providecommand \urlprefix  [0]{URL }%
	\providecommand \Eprint [0]{\href }%
	\providecommand \doibase [0]{https://doi.org/}%
	\providecommand \selectlanguage [0]{\@gobble}%
	\providecommand \bibinfo  [0]{\@secondoftwo}%
	\providecommand \bibfield  [0]{\@secondoftwo}%
	\providecommand \translation [1]{[#1]}%
	\providecommand \BibitemOpen [0]{}%
	\providecommand \bibitemStop [0]{}%
	\providecommand \bibitemNoStop [0]{.\EOS\space}%
	\providecommand \EOS [0]{\spacefactor3000\relax}%
	\providecommand \BibitemShut  [1]{\csname bibitem#1\endcsname}%
	\let\auto@bib@innerbib\@empty
	\bibitem [{\citenamefont {Massignan}\ \emph {et~al.}(2014)\citenamefont
		{Massignan}, \citenamefont {Zaccanti},\ and\ \citenamefont
		{Bruun}}]{2014_Massignan_RepProgPhys_Review}%
	\BibitemOpen
	\bibfield  {author} {\bibinfo {author} {\bibfnamefont {P.}~\bibnamefont
			{Massignan}}, \bibinfo {author} {\bibfnamefont {M.}~\bibnamefont
			{Zaccanti}},\ and\ \bibinfo {author} {\bibfnamefont {G.~M.}\ \bibnamefont
			{Bruun}},\ }\bibfield  {title} {\bibinfo {title} {Polarons, dressed molecules
			and itinerant ferromagnetism in ultracold fermi gases},\ }\href
	{https://doi.org/10.1088/0034-4885/77/3/034401} {\bibfield  {journal}
		{\bibinfo  {journal} {Rep Prog Phys}\ }\textbf {\bibinfo {volume} {77}},\
		\bibinfo {pages} {034401} (\bibinfo {year} {2014})}\BibitemShut {NoStop}%
	\bibitem [{\citenamefont {Levinsen}\ and\ \citenamefont
		{Parish}()}]{FermiGasesReview}%
	\BibitemOpen
	\bibfield  {author} {\bibinfo {author} {\bibfnamefont {J.}~\bibnamefont
			{Levinsen}}\ and\ \bibinfo {author} {\bibfnamefont {M.~M.}\ \bibnamefont
			{Parish}},\ }\bibinfo {title} {Strongly interacting two-dimensional fermi
		gases},\ in\ \href {https://doi.org/10.1142/9789814667746_0001} {\emph
		{\bibinfo {booktitle} {Annual Review of Cold Atoms and Molecules}}},\
	Chap.~\bibinfo {chapter} {1}, pp.\ \bibinfo {pages} {1--75}\BibitemShut
	{NoStop}%
	\bibitem [{\citenamefont {Cetina}\ \emph {et~al.}(2016)\citenamefont {Cetina},
		\citenamefont {Jag}, \citenamefont {Lous}, \citenamefont {Fritsche},
		\citenamefont {Walraven}, \citenamefont {Grimm}, \citenamefont {Levinsen},
		\citenamefont {Parish}, \citenamefont {Schmidt}, \citenamefont {Knap},\ and\
		\citenamefont {Demler}}]{Cetina2016}%
	\BibitemOpen
	\bibfield  {author} {\bibinfo {author} {\bibfnamefont {M.}~\bibnamefont
			{Cetina}}, \bibinfo {author} {\bibfnamefont {M.}~\bibnamefont {Jag}},
		\bibinfo {author} {\bibfnamefont {R.~S.}\ \bibnamefont {Lous}}, \bibinfo
		{author} {\bibfnamefont {I.}~\bibnamefont {Fritsche}}, \bibinfo {author}
		{\bibfnamefont {J.~T.~M.}\ \bibnamefont {Walraven}}, \bibinfo {author}
		{\bibfnamefont {R.}~\bibnamefont {Grimm}}, \bibinfo {author} {\bibfnamefont
			{J.}~\bibnamefont {Levinsen}}, \bibinfo {author} {\bibfnamefont {M.~M.}\
			\bibnamefont {Parish}}, \bibinfo {author} {\bibfnamefont {R.}~\bibnamefont
			{Schmidt}}, \bibinfo {author} {\bibfnamefont {M.}~\bibnamefont {Knap}},\ and\
		\bibinfo {author} {\bibfnamefont {E.}~\bibnamefont {Demler}},\ }\bibfield
	{title} {\bibinfo {title} {{Ultrafast many-body interferometry of impurities
				coupled to a Fermi sea}},\ }\href {https://doi.org/10.1126/science.aaf5134}
	{\bibfield  {journal} {\bibinfo  {journal} {Science}\ }\textbf {\bibinfo
			{volume} {354}},\ \bibinfo {pages} {96} (\bibinfo {year} {2016})}\BibitemShut
	{NoStop}%
	\bibitem [{\citenamefont {Schirotzek}\ \emph {et~al.}(2009)\citenamefont
		{Schirotzek}, \citenamefont {Wu}, \citenamefont {Sommer},\ and\ \citenamefont
		{Zwierlein}}]{Schirotzek2009}%
	\BibitemOpen
	\bibfield  {author} {\bibinfo {author} {\bibfnamefont {A.}~\bibnamefont
			{Schirotzek}}, \bibinfo {author} {\bibfnamefont {C.-H.}\ \bibnamefont {Wu}},
		\bibinfo {author} {\bibfnamefont {A.}~\bibnamefont {Sommer}},\ and\ \bibinfo
		{author} {\bibfnamefont {M.~W.}\ \bibnamefont {Zwierlein}},\ }\bibfield
	{title} {\bibinfo {title} {{Observation of Fermi Polarons in a Tunable Fermi
				Liquid of Ultracold Atoms}},\ }\href
	{https://doi.org/10.1103/PhysRevLett.102.230402} {\bibfield  {journal}
		{\bibinfo  {journal} {Phys. Rev. Lett.}\ }\textbf {\bibinfo {volume} {102}},\
		\bibinfo {pages} {230402} (\bibinfo {year} {2009})}\BibitemShut {NoStop}%
	\bibitem [{\citenamefont {Kohstall}\ \emph {et~al.}(2012)\citenamefont
		{Kohstall}, \citenamefont {Zaccanti}, \citenamefont {Jag}, \citenamefont
		{Trenkwalder}, \citenamefont {Massignan}, \citenamefont {Bruun},
		\citenamefont {Schreck},\ and\ \citenamefont
		{Grimm}}]{2012_C.Kohstall_Nature_FermiPolColdAtom}%
	\BibitemOpen
	\bibfield  {author} {\bibinfo {author} {\bibfnamefont {C.}~\bibnamefont
			{Kohstall}}, \bibinfo {author} {\bibfnamefont {M.}~\bibnamefont {Zaccanti}},
		\bibinfo {author} {\bibfnamefont {M.}~\bibnamefont {Jag}}, \bibinfo {author}
		{\bibfnamefont {A.}~\bibnamefont {Trenkwalder}}, \bibinfo {author}
		{\bibfnamefont {P.}~\bibnamefont {Massignan}}, \bibinfo {author}
		{\bibfnamefont {G.~M.}\ \bibnamefont {Bruun}}, \bibinfo {author}
		{\bibfnamefont {F.}~\bibnamefont {Schreck}},\ and\ \bibinfo {author}
		{\bibfnamefont {R.}~\bibnamefont {Grimm}},\ }\bibfield  {title} {\bibinfo
		{title} {Metastability and coherence of repulsive polarons in a strongly
			interacting fermi mixture},\ }\href {https://doi.org/10.1038/nature11065}
	{\bibfield  {journal} {\bibinfo  {journal} {Nature}\ }\textbf {\bibinfo
			{volume} {485}},\ \bibinfo {pages} {615} (\bibinfo {year}
		{2012})}\BibitemShut {NoStop}%
	\bibitem [{\citenamefont {Nascimbène}\ \emph {et~al.}(2009)\citenamefont
		{Nascimbène}, \citenamefont {Navon}, \citenamefont {Jiang}, \citenamefont
		{Tarruell}, \citenamefont {Teichmann}, \citenamefont {McKeever},
		\citenamefont {Chevy},\ and\ \citenamefont
		{Salomon}}]{2009_Salomon_PRL_3DFermiPolaron}%
	\BibitemOpen
	\bibfield  {author} {\bibinfo {author} {\bibfnamefont {S.}~\bibnamefont
			{Nascimbène}}, \bibinfo {author} {\bibfnamefont {N.}~\bibnamefont {Navon}},
		\bibinfo {author} {\bibfnamefont {K.~J.}\ \bibnamefont {Jiang}}, \bibinfo
		{author} {\bibfnamefont {L.}~\bibnamefont {Tarruell}}, \bibinfo {author}
		{\bibfnamefont {M.}~\bibnamefont {Teichmann}}, \bibinfo {author}
		{\bibfnamefont {J.}~\bibnamefont {McKeever}}, \bibinfo {author}
		{\bibfnamefont {F.}~\bibnamefont {Chevy}},\ and\ \bibinfo {author}
		{\bibfnamefont {C.}~\bibnamefont {Salomon}},\ }\bibfield  {title} {\bibinfo
		{title} {Collective oscillations of an imbalanced fermi gas: Axial
			compression modes and polaron effective mass},\ }\href
	{https://doi.org/10.1103/physrevlett.103.170402} {\bibfield  {journal}
		{\bibinfo  {journal} {Physical Review Letters}\ }\textbf {\bibinfo {volume}
			{103}},\ \bibinfo {pages} {170402} (\bibinfo {year} {2009})}\BibitemShut
	{NoStop}%
	\bibitem [{\citenamefont {McGuire}(1966)}]{McGuire1966}%
	\BibitemOpen
	\bibfield  {author} {\bibinfo {author} {\bibfnamefont {J.~B.}\ \bibnamefont
			{McGuire}},\ }\bibfield  {title} {\bibinfo {title} {Interacting fermions in
			one dimension. ii. attractive potential},\ }\href
	{https://doi.org/10.1063/1.1704798} {\bibfield  {journal} {\bibinfo
			{journal} {Journal of Mathematical Physics}\ }\textbf {\bibinfo {volume}
			{7}},\ \bibinfo {pages} {123} (\bibinfo {year} {1966})}\BibitemShut {NoStop}%
	\bibitem [{\citenamefont {Sidler}\ \emph {et~al.}(2017)\citenamefont {Sidler},
		\citenamefont {Back}, \citenamefont {Cotlet}, \citenamefont {Srivastava},
		\citenamefont {Fink}, \citenamefont {Kroner}, \citenamefont {Demler},\ and\
		\citenamefont {Imamoglu}}]{2016_A.Imamoglu_NatPhys_MoSe2}%
	\BibitemOpen
	\bibfield  {author} {\bibinfo {author} {\bibfnamefont {M.}~\bibnamefont
			{Sidler}}, \bibinfo {author} {\bibfnamefont {P.}~\bibnamefont {Back}},
		\bibinfo {author} {\bibfnamefont {O.}~\bibnamefont {Cotlet}}, \bibinfo
		{author} {\bibfnamefont {A.}~\bibnamefont {Srivastava}}, \bibinfo {author}
		{\bibfnamefont {T.}~\bibnamefont {Fink}}, \bibinfo {author} {\bibfnamefont
			{M.}~\bibnamefont {Kroner}}, \bibinfo {author} {\bibfnamefont
			{E.}~\bibnamefont {Demler}},\ and\ \bibinfo {author} {\bibfnamefont
			{A.}~\bibnamefont {Imamoglu}},\ }\bibfield  {title} {\bibinfo {title} {Fermi
			polaron-polaritons in charge-tunable atomically thin semiconductors},\ }\href
	{https://doi.org/10.1038/nphys3949} {\bibfield  {journal} {\bibinfo
			{journal} {Nature Physics}\ }\textbf {\bibinfo {volume} {13}},\ \bibinfo
		{pages} {255} (\bibinfo {year} {2017})}\BibitemShut {NoStop}%
	\bibitem [{\citenamefont {Efimkin}\ and\ \citenamefont
		{Macdonald}(2017)}]{2017_Dmitry_PRB_theory}%
	\BibitemOpen
	\bibfield  {author} {\bibinfo {author} {\bibfnamefont {D.~K.}\ \bibnamefont
			{Efimkin}}\ and\ \bibinfo {author} {\bibfnamefont {A.~H.}\ \bibnamefont
			{Macdonald}},\ }\bibfield  {title} {\bibinfo {title} {Many-body theory of
			trion absorption features in two-dimensional semiconductors},\ }\href
	{https://doi.org/10.1103/physrevb.95.035417} {\bibfield  {journal} {\bibinfo
			{journal} {Physical Review B}\ }\textbf {\bibinfo {volume} {95}},\ \bibinfo
		{pages} {035417} (\bibinfo {year} {2017})}\BibitemShut {NoStop}%
	\bibitem [{\citenamefont {Mueller}\ and\ \citenamefont
		{Malic}(2018)}]{2018_M.Ermin_TMD_Exciton_Review}%
	\BibitemOpen
	\bibfield  {author} {\bibinfo {author} {\bibfnamefont {T.}~\bibnamefont
			{Mueller}}\ and\ \bibinfo {author} {\bibfnamefont {E.}~\bibnamefont
			{Malic}},\ }\bibfield  {title} {\bibinfo {title} {Exciton physics and device
			application of two-dimensional transition metal dichalcogenide
			semiconductors},\ }\href {https://doi.org/10.1038/s41699-018-0074-2}
	{\bibfield  {journal} {\bibinfo  {journal} {npj 2D Materials and
				Applications}\ }\textbf {\bibinfo {volume} {2}},\ \bibinfo {pages} {29}
		(\bibinfo {year} {2018})}\BibitemShut {NoStop}%
	\bibitem [{\citenamefont {He}\ \emph {et~al.}(2020)\citenamefont {He},
		\citenamefont {Rivera}, \citenamefont {Van~Tuan}, \citenamefont {Wilson},
		\citenamefont {Yang}, \citenamefont {Taniguchi}, \citenamefont {Watanabe},
		\citenamefont {Yan}, \citenamefont {Mandrus}, \citenamefont {Yu},
		\citenamefont {Dery}, \citenamefont {Yao},\ and\ \citenamefont
		{Xu}}]{2020_He_NatComm_ExcitonComplexes}%
	\BibitemOpen
	\bibfield  {author} {\bibinfo {author} {\bibfnamefont {M.}~\bibnamefont
			{He}}, \bibinfo {author} {\bibfnamefont {P.}~\bibnamefont {Rivera}}, \bibinfo
		{author} {\bibfnamefont {D.}~\bibnamefont {Van~Tuan}}, \bibinfo {author}
		{\bibfnamefont {N.~P.}\ \bibnamefont {Wilson}}, \bibinfo {author}
		{\bibfnamefont {M.}~\bibnamefont {Yang}}, \bibinfo {author} {\bibfnamefont
			{T.}~\bibnamefont {Taniguchi}}, \bibinfo {author} {\bibfnamefont
			{K.}~\bibnamefont {Watanabe}}, \bibinfo {author} {\bibfnamefont
			{J.}~\bibnamefont {Yan}}, \bibinfo {author} {\bibfnamefont {D.~G.}\
			\bibnamefont {Mandrus}}, \bibinfo {author} {\bibfnamefont {H.}~\bibnamefont
			{Yu}}, \bibinfo {author} {\bibfnamefont {H.}~\bibnamefont {Dery}}, \bibinfo
		{author} {\bibfnamefont {W.}~\bibnamefont {Yao}},\ and\ \bibinfo {author}
		{\bibfnamefont {X.}~\bibnamefont {Xu}},\ }\bibfield  {title} {\bibinfo
		{title} {Valley phonons and exciton complexes in a monolayer semiconductor},\
	}\bibfield  {journal} {\bibinfo  {journal} {Nature Communications}\ }\textbf
	{\bibinfo {volume} {11}},\ \href {https://doi.org/10.1038/s41467-020-14472-0}
	{10.1038/s41467-020-14472-0} (\bibinfo {year} {2020})\BibitemShut {NoStop}%
	\bibitem [{\citenamefont {Mak}\ \emph {et~al.}(2013)\citenamefont {Mak},
		\citenamefont {He}, \citenamefont {Lee}, \citenamefont {Lee}, \citenamefont
		{Hone}, \citenamefont {Heinz},\ and\ \citenamefont
		{Shan}}]{2013_KF.Mak_Nat.Mat._Trion}%
	\BibitemOpen
	\bibfield  {author} {\bibinfo {author} {\bibfnamefont {K.~F.}\ \bibnamefont
			{Mak}}, \bibinfo {author} {\bibfnamefont {K.}~\bibnamefont {He}}, \bibinfo
		{author} {\bibfnamefont {C.}~\bibnamefont {Lee}}, \bibinfo {author}
		{\bibfnamefont {G.~H.}\ \bibnamefont {Lee}}, \bibinfo {author} {\bibfnamefont
			{J.}~\bibnamefont {Hone}}, \bibinfo {author} {\bibfnamefont {T.~F.}\
			\bibnamefont {Heinz}},\ and\ \bibinfo {author} {\bibfnamefont
			{J.}~\bibnamefont {Shan}},\ }\bibfield  {title} {\bibinfo {title} {Tightly
			bound trions in monolayer mos$_2$},\ }\href
	{https://doi.org/10.1038/nmat3505} {\bibfield  {journal} {\bibinfo  {journal}
			{Nature Materials}\ }\textbf {\bibinfo {volume} {12}},\ \bibinfo {pages}
		{207} (\bibinfo {year} {2013})}\BibitemShut {NoStop}%
	\bibitem [{\citenamefont {Glazov}(2020)}]{2020_Glazov_JCP_EqualTrionPolaron}%
	\BibitemOpen
	\bibfield  {author} {\bibinfo {author} {\bibfnamefont {M.~M.}\ \bibnamefont
			{Glazov}},\ }\bibfield  {title} {\bibinfo {title} {Optical properties of
			charged excitons in two-dimensional semiconductors},\ }\href
	{https://doi.org/10.1063/5.0012475} {\bibfield  {journal} {\bibinfo
			{journal} {The Journal of Chemical Physics}\ }\textbf {\bibinfo {volume}
			{153}},\ \bibinfo {pages} {034703} (\bibinfo {year} {2020})}\BibitemShut
	{NoStop}%
	\bibitem [{\citenamefont {Suris}\ \emph {et~al.}(2001)\citenamefont {Suris},
		\citenamefont {Kochereshko}, \citenamefont {Astakhov}, \citenamefont
		{Yakovlev}, \citenamefont {Ossau}, \citenamefont {Nürnberger}, \citenamefont
		{Faschinger}, \citenamefont {Landwehr}, \citenamefont {Wojtowicz},
		\citenamefont {Karczewski},\ and\ \citenamefont {Kossut}}]{TrionDoubtSuris}%
	\BibitemOpen
	\bibfield  {author} {\bibinfo {author} {\bibfnamefont {R.}~\bibnamefont
			{Suris}}, \bibinfo {author} {\bibfnamefont {V.}~\bibnamefont {Kochereshko}},
		\bibinfo {author} {\bibfnamefont {G.}~\bibnamefont {Astakhov}}, \bibinfo
		{author} {\bibfnamefont {D.}~\bibnamefont {Yakovlev}}, \bibinfo {author}
		{\bibfnamefont {W.}~\bibnamefont {Ossau}}, \bibinfo {author} {\bibfnamefont
			{J.}~\bibnamefont {Nürnberger}}, \bibinfo {author} {\bibfnamefont
			{W.}~\bibnamefont {Faschinger}}, \bibinfo {author} {\bibfnamefont
			{G.}~\bibnamefont {Landwehr}}, \bibinfo {author} {\bibfnamefont
			{T.}~\bibnamefont {Wojtowicz}}, \bibinfo {author} {\bibfnamefont
			{G.}~\bibnamefont {Karczewski}},\ and\ \bibinfo {author} {\bibfnamefont
			{J.}~\bibnamefont {Kossut}},\ }\bibfield  {title} {\bibinfo {title} {Excitons
			and trions modified by interaction with a two-dimensional electron gas},\
	}\href
	{https://doi.org/https://doi.org/10.1002/1521-3951(200110)227:2<343::AID-PSSB343>3.0.CO;2-W}
	{\bibfield  {journal} {\bibinfo  {journal} {physica status solidi (b)}\
		}\textbf {\bibinfo {volume} {227}},\ \bibinfo {pages} {343} (\bibinfo {year}
		{2001})}\BibitemShut {NoStop}%
	\bibitem [{\citenamefont {Shiau}\ \emph {et~al.}(2012)\citenamefont {Shiau},
		\citenamefont {Combescot},\ and\ \citenamefont
		{Chang}}]{TrionDoubtCombescot}%
	\BibitemOpen
	\bibfield  {author} {\bibinfo {author} {\bibfnamefont {S.-Y.}\ \bibnamefont
			{Shiau}}, \bibinfo {author} {\bibfnamefont {M.}~\bibnamefont {Combescot}},\
		and\ \bibinfo {author} {\bibfnamefont {Y.-C.}\ \bibnamefont {Chang}},\
	}\bibfield  {title} {\bibinfo {title} {Trion ground state, excited states,
			and absorption spectrum using electron-exciton basis},\ }\href
	{https://doi.org/10.1103/PhysRevB.86.115210} {\bibfield  {journal} {\bibinfo
			{journal} {Phys. Rev. B}\ }\textbf {\bibinfo {volume} {86}},\ \bibinfo
		{pages} {115210} (\bibinfo {year} {2012})}\BibitemShut {NoStop}%
	\bibitem [{\citenamefont {Baeten}\ and\ \citenamefont
		{Wouters}(2015)}]{TrionDoubtWouters2}%
	\BibitemOpen
	\bibfield  {author} {\bibinfo {author} {\bibfnamefont {M.}~\bibnamefont
			{Baeten}}\ and\ \bibinfo {author} {\bibfnamefont {M.}~\bibnamefont
			{Wouters}},\ }\bibfield  {title} {\bibinfo {title} {Many-body effects of a
			two-dimensional electron gas on trion-polaritons},\ }\href
	{https://doi.org/10.1103/PhysRevB.91.115313} {\bibfield  {journal} {\bibinfo
			{journal} {Phys. Rev. B}\ }\textbf {\bibinfo {volume} {91}},\ \bibinfo
		{pages} {115313} (\bibinfo {year} {2015})}\BibitemShut {NoStop}%
	\bibitem [{\citenamefont {Li}\ \emph {et~al.}(2022)\citenamefont {Li},
		\citenamefont {Goryca}, \citenamefont {Choi}, \citenamefont {Xu},\ and\
		\citenamefont {Crooker}}]{2022_S.Crooker_NanoLett_WSe2ManyBody}%
	\BibitemOpen
	\bibfield  {author} {\bibinfo {author} {\bibfnamefont {J.}~\bibnamefont
			{Li}}, \bibinfo {author} {\bibfnamefont {M.}~\bibnamefont {Goryca}}, \bibinfo
		{author} {\bibfnamefont {J.}~\bibnamefont {Choi}}, \bibinfo {author}
		{\bibfnamefont {X.}~\bibnamefont {Xu}},\ and\ \bibinfo {author}
		{\bibfnamefont {S.~A.}\ \bibnamefont {Crooker}},\ }\bibfield  {title}
	{\bibinfo {title} {Many-body exciton and intervalley correlations in heavily
			electron-doped wse2 monolayers},\ }\href
	{https://doi.org/10.1021/acs.nanolett.1c04217} {\bibfield  {journal}
		{\bibinfo  {journal} {Nano Letters}\ }\textbf {\bibinfo {volume} {22}},\
		\bibinfo {pages} {426} (\bibinfo {year} {2022})}\BibitemShut {NoStop}%
	\bibitem [{\citenamefont {Wagner}\ \emph {et~al.}(2020)\citenamefont {Wagner},
		\citenamefont {Wietek}, \citenamefont {Ziegler}, \citenamefont {Semina},
		\citenamefont {Taniguchi}, \citenamefont {Watanabe}, \citenamefont {Zipfel},
		\citenamefont {Glazov},\ and\ \citenamefont
		{Chernikov}}]{2020_A.Chernikov_PRL_WSe2Trion}%
	\BibitemOpen
	\bibfield  {author} {\bibinfo {author} {\bibfnamefont {K.}~\bibnamefont
			{Wagner}}, \bibinfo {author} {\bibfnamefont {E.}~\bibnamefont {Wietek}},
		\bibinfo {author} {\bibfnamefont {J.~D.}\ \bibnamefont {Ziegler}}, \bibinfo
		{author} {\bibfnamefont {M.~A.}\ \bibnamefont {Semina}}, \bibinfo {author}
		{\bibfnamefont {T.}~\bibnamefont {Taniguchi}}, \bibinfo {author}
		{\bibfnamefont {K.}~\bibnamefont {Watanabe}}, \bibinfo {author}
		{\bibfnamefont {J.}~\bibnamefont {Zipfel}}, \bibinfo {author} {\bibfnamefont
			{M.~M.}\ \bibnamefont {Glazov}},\ and\ \bibinfo {author} {\bibfnamefont
			{A.}~\bibnamefont {Chernikov}},\ }\bibfield  {title} {\bibinfo {title}
		{{Autoionization and Dressing of Excited Excitons by Free Carriers in
				Monolayer WSe$_2$}},\ }\href {https://doi.org/10.1103/physrevlett.125.267401}
	{\bibfield  {journal} {\bibinfo  {journal} {Physical Review Letters}\
		}\textbf {\bibinfo {volume} {125}},\ \bibinfo {pages} {267401} (\bibinfo
		{year} {2020})}\BibitemShut {NoStop}%
	\bibitem [{\citenamefont {Van~Tuan}\ \emph {et~al.}(2022)\citenamefont
		{Van~Tuan}, \citenamefont {Shi}, \citenamefont {Xu}, \citenamefont
		{Crooker},\ and\ \citenamefont {Dery}}]{2022_S.Crooker_hexcitons&oxciton}%
	\BibitemOpen
	\bibfield  {author} {\bibinfo {author} {\bibfnamefont {D.}~\bibnamefont
			{Van~Tuan}}, \bibinfo {author} {\bibfnamefont {S.-F.}\ \bibnamefont {Shi}},
		\bibinfo {author} {\bibfnamefont {X.}~\bibnamefont {Xu}}, \bibinfo {author}
		{\bibfnamefont {S.~A.}\ \bibnamefont {Crooker}},\ and\ \bibinfo {author}
		{\bibfnamefont {H.}~\bibnamefont {Dery}},\ }\bibfield  {title} {\bibinfo
		{title} {{Hexcitons and oxcitons in monolayer WSe$_2$}},\ }\href@noop {}
	{\bibfield  {journal} {\bibinfo  {journal} {arXiv preprint:2202.08375}\ }
		(\bibinfo {year} {2022})}\BibitemShut {NoStop}%
	\bibitem [{\citenamefont {Smoleński}\ \emph {et~al.}(2019)\citenamefont
		{Smoleński}, \citenamefont {Cotlet}, \citenamefont {Popert}, \citenamefont
		{Back}, \citenamefont {Shimazaki}, \citenamefont {Knüppel}, \citenamefont
		{Dietler}, \citenamefont {Taniguchi}, \citenamefont {Watanabe}, \citenamefont
		{Kroner},\ and\ \citenamefont
		{Imamoglu}}]{2019_A.Imamoglu_PRL_PolaronMagField}%
	\BibitemOpen
	\bibfield  {author} {\bibinfo {author} {\bibfnamefont {T.}~\bibnamefont
			{Smoleński}}, \bibinfo {author} {\bibfnamefont {O.}~\bibnamefont {Cotlet}},
		\bibinfo {author} {\bibfnamefont {A.}~\bibnamefont {Popert}}, \bibinfo
		{author} {\bibfnamefont {P.}~\bibnamefont {Back}}, \bibinfo {author}
		{\bibfnamefont {Y.}~\bibnamefont {Shimazaki}}, \bibinfo {author}
		{\bibfnamefont {P.}~\bibnamefont {Knüppel}}, \bibinfo {author}
		{\bibfnamefont {N.}~\bibnamefont {Dietler}}, \bibinfo {author} {\bibfnamefont
			{T.}~\bibnamefont {Taniguchi}}, \bibinfo {author} {\bibfnamefont
			{K.}~\bibnamefont {Watanabe}}, \bibinfo {author} {\bibfnamefont
			{M.}~\bibnamefont {Kroner}},\ and\ \bibinfo {author} {\bibfnamefont
			{A.}~\bibnamefont {Imamoglu}},\ }\bibfield  {title} {\bibinfo {title}
		{{Interaction-Induced Shubnikov–de Haas Oscillations in Optical
				Conductivity of Monolayer MoSe$_2$}},\ }\href
	{https://doi.org/10.1103/physrevlett.123.097403} {\bibfield  {journal}
		{\bibinfo  {journal} {Physical Review Letters}\ }\textbf {\bibinfo {volume}
			{123}},\ \bibinfo {pages} {097403} (\bibinfo {year} {2019})}\BibitemShut
	{NoStop}%
	\bibitem [{\citenamefont {Siemens}\ \emph {et~al.}(2010)\citenamefont
		{Siemens}, \citenamefont {Moody}, \citenamefont {Li}, \citenamefont
		{Bristow},\ and\ \citenamefont {Cundiff}}]{2010_Mark_OE_2DCSLineShape}%
	\BibitemOpen
	\bibfield  {author} {\bibinfo {author} {\bibfnamefont {M.~E.}\ \bibnamefont
			{Siemens}}, \bibinfo {author} {\bibfnamefont {G.}~\bibnamefont {Moody}},
		\bibinfo {author} {\bibfnamefont {H.}~\bibnamefont {Li}}, \bibinfo {author}
		{\bibfnamefont {A.~D.}\ \bibnamefont {Bristow}},\ and\ \bibinfo {author}
		{\bibfnamefont {S.~T.}\ \bibnamefont {Cundiff}},\ }\bibfield  {title}
	{\bibinfo {title} {Resonance lineshapes in two-dimensional fourier transform
			spectroscopy},\ }\href {https://doi.org/10.1364/OE.18.017699} {\bibfield
		{journal} {\bibinfo  {journal} {Opt Express}\ }\textbf {\bibinfo {volume}
			{18}},\ \bibinfo {pages} {17699} (\bibinfo {year} {2010})}\BibitemShut
	{NoStop}%
	\bibitem [{\citenamefont {Moody}\ \emph {et~al.}(2015)\citenamefont {Moody},
		\citenamefont {Kavir~Dass}, \citenamefont {Hao}, \citenamefont {Chen},
		\citenamefont {Li}, \citenamefont {Singh}, \citenamefont {Tran},
		\citenamefont {Clark}, \citenamefont {Xu}, \citenamefont {Berghauser},
		\citenamefont {Malic}, \citenamefont {Knorr},\ and\ \citenamefont
		{Li}}]{2015_G.Moody_NatComm_2DCSMoSe2}%
	\BibitemOpen
	\bibfield  {author} {\bibinfo {author} {\bibfnamefont {G.}~\bibnamefont
			{Moody}}, \bibinfo {author} {\bibfnamefont {C.}~\bibnamefont {Kavir~Dass}},
		\bibinfo {author} {\bibfnamefont {K.}~\bibnamefont {Hao}}, \bibinfo {author}
		{\bibfnamefont {C.~H.}\ \bibnamefont {Chen}}, \bibinfo {author}
		{\bibfnamefont {L.~J.}\ \bibnamefont {Li}}, \bibinfo {author} {\bibfnamefont
			{A.}~\bibnamefont {Singh}}, \bibinfo {author} {\bibfnamefont
			{K.}~\bibnamefont {Tran}}, \bibinfo {author} {\bibfnamefont {G.}~\bibnamefont
			{Clark}}, \bibinfo {author} {\bibfnamefont {X.}~\bibnamefont {Xu}}, \bibinfo
		{author} {\bibfnamefont {G.}~\bibnamefont {Berghauser}}, \bibinfo {author}
		{\bibfnamefont {E.}~\bibnamefont {Malic}}, \bibinfo {author} {\bibfnamefont
			{A.}~\bibnamefont {Knorr}},\ and\ \bibinfo {author} {\bibfnamefont
			{X.}~\bibnamefont {Li}},\ }\bibfield  {title} {\bibinfo {title} {Intrinsic
			homogeneous linewidth and broadening mechanisms of excitons in monolayer
			transition metal dichalcogenides},\ }\href
	{https://doi.org/10.1038/ncomms9315} {\bibfield  {journal} {\bibinfo
			{journal} {Nature Communications}\ }\textbf {\bibinfo {volume} {6}},\
		\bibinfo {pages} {8315} (\bibinfo {year} {2015})}\BibitemShut {NoStop}%
	\bibitem [{\citenamefont {Martin}\ \emph {et~al.}(2020)\citenamefont {Martin},
		\citenamefont {Horng}, \citenamefont {Ruth}, \citenamefont {Paik},
		\citenamefont {Wentzel}, \citenamefont {Deng},\ and\ \citenamefont
		{Cundiff}}]{2020_E.Martin_PRApp_2DCSMoSe2}%
	\BibitemOpen
	\bibfield  {author} {\bibinfo {author} {\bibfnamefont {E.~W.}\ \bibnamefont
			{Martin}}, \bibinfo {author} {\bibfnamefont {J.}~\bibnamefont {Horng}},
		\bibinfo {author} {\bibfnamefont {H.~G.}\ \bibnamefont {Ruth}}, \bibinfo
		{author} {\bibfnamefont {E.}~\bibnamefont {Paik}}, \bibinfo {author}
		{\bibfnamefont {M.-H.}\ \bibnamefont {Wentzel}}, \bibinfo {author}
		{\bibfnamefont {H.}~\bibnamefont {Deng}},\ and\ \bibinfo {author}
		{\bibfnamefont {S.~T.}\ \bibnamefont {Cundiff}},\ }\bibfield  {title}
	{\bibinfo {title} {{Encapsulation Narrows and Preserves the Excitonic
				Homogeneous Linewidth of Exfoliated Monolayer MoSe$_2$}},\ }\href
	{https://doi.org/10.1103/physrevapplied.14.021002} {\bibfield  {journal}
		{\bibinfo  {journal} {Physical Review Applied}\ }\textbf {\bibinfo {volume}
			{14}},\ \bibinfo {pages} {021002} (\bibinfo {year} {2020})}\BibitemShut
	{NoStop}%
	\bibitem [{\citenamefont {Hao}\ \emph {et~al.}(2016)\citenamefont {Hao},
		\citenamefont {Xu}, \citenamefont {Nagler}, \citenamefont {Singh},
		\citenamefont {Tran}, \citenamefont {Dass}, \citenamefont {Schüller},
		\citenamefont {Korn}, \citenamefont {Li},\ and\ \citenamefont
		{Moody}}]{2016_H.Kai_NanoLett_CP-2DCS}%
	\BibitemOpen
	\bibfield  {author} {\bibinfo {author} {\bibfnamefont {K.}~\bibnamefont
			{Hao}}, \bibinfo {author} {\bibfnamefont {L.}~\bibnamefont {Xu}}, \bibinfo
		{author} {\bibfnamefont {P.}~\bibnamefont {Nagler}}, \bibinfo {author}
		{\bibfnamefont {A.}~\bibnamefont {Singh}}, \bibinfo {author} {\bibfnamefont
			{K.}~\bibnamefont {Tran}}, \bibinfo {author} {\bibfnamefont {C.~K.}\
			\bibnamefont {Dass}}, \bibinfo {author} {\bibfnamefont {C.}~\bibnamefont
			{Schüller}}, \bibinfo {author} {\bibfnamefont {T.}~\bibnamefont {Korn}},
		\bibinfo {author} {\bibfnamefont {X.}~\bibnamefont {Li}},\ and\ \bibinfo
		{author} {\bibfnamefont {G.}~\bibnamefont {Moody}},\ }\bibfield  {title}
	{\bibinfo {title} {Coherent and incoherent coupling dynamics between neutral
			and charged excitons in monolayer mose$_2$},\ }\href
	{https://doi.org/10.1021/acs.nanolett.6b02041} {\bibfield  {journal}
		{\bibinfo  {journal} {Nano Letters}\ }\textbf {\bibinfo {volume} {16}},\
		\bibinfo {pages} {5109} (\bibinfo {year} {2016})}\BibitemShut {NoStop}%
	\bibitem [{\citenamefont {Liu}\ \emph {et~al.}(2021)\citenamefont {Liu},
		\citenamefont {Van~Baren}, \citenamefont {Lu}, \citenamefont {Taniguchi},
		\citenamefont {Watanabe}, \citenamefont {Smirnov}, \citenamefont {Chang},\
		and\ \citenamefont {Lui}}]{2021_CHLui_NatComm_MoSe2&WSe2}%
	\BibitemOpen
	\bibfield  {author} {\bibinfo {author} {\bibfnamefont {E.}~\bibnamefont
			{Liu}}, \bibinfo {author} {\bibfnamefont {J.}~\bibnamefont {Van~Baren}},
		\bibinfo {author} {\bibfnamefont {Z.}~\bibnamefont {Lu}}, \bibinfo {author}
		{\bibfnamefont {T.}~\bibnamefont {Taniguchi}}, \bibinfo {author}
		{\bibfnamefont {K.}~\bibnamefont {Watanabe}}, \bibinfo {author}
		{\bibfnamefont {D.}~\bibnamefont {Smirnov}}, \bibinfo {author} {\bibfnamefont
			{Y.-C.}\ \bibnamefont {Chang}},\ and\ \bibinfo {author} {\bibfnamefont
			{C.~H.}\ \bibnamefont {Lui}},\ }\bibfield  {title} {\bibinfo {title}
		{Exciton-polaron rydberg states in monolayer mose2 and wse2},\ }\bibfield
	{journal} {\bibinfo  {journal} {Nature Communications}\ }\textbf {\bibinfo
		{volume} {12}},\ \href {https://doi.org/10.1038/s41467-021-26304-w}
	{10.1038/s41467-021-26304-w} (\bibinfo {year} {2021})\BibitemShut {NoStop}%
	\bibitem [{\citenamefont {Chevy}(2006)}]{Chevy2006}%
	\BibitemOpen
	\bibfield  {author} {\bibinfo {author} {\bibfnamefont {F.}~\bibnamefont
			{Chevy}},\ }\bibfield  {title} {\bibinfo {title} {Universal phase diagram of
			a strongly interacting fermi gas with unbalanced spin populations},\ }\href
	{https://doi.org/10.1103/PhysRevA.74.063628} {\bibfield  {journal} {\bibinfo
			{journal} {Phys. Rev. A}\ }\textbf {\bibinfo {volume} {74}},\ \bibinfo
		{pages} {063628} (\bibinfo {year} {2006})}\BibitemShut {NoStop}%
	\bibitem [{\citenamefont {Tiene}\ \emph {et~al.}(2022)\citenamefont {Tiene},
		\citenamefont {Levinsen}, \citenamefont {Keeling}, \citenamefont {Parish},\
		and\ \citenamefont {Marchetti}}]{Tiene2022}%
	\BibitemOpen
	\bibfield  {author} {\bibinfo {author} {\bibfnamefont {A.}~\bibnamefont
			{Tiene}}, \bibinfo {author} {\bibfnamefont {J.}~\bibnamefont {Levinsen}},
		\bibinfo {author} {\bibfnamefont {J.}~\bibnamefont {Keeling}}, \bibinfo
		{author} {\bibfnamefont {M.~M.}\ \bibnamefont {Parish}},\ and\ \bibinfo
		{author} {\bibfnamefont {F.~M.}\ \bibnamefont {Marchetti}},\ }\bibfield
	{title} {\bibinfo {title} {Effect of fermion indistinguishability on optical
			absorption of doped two-dimensional semiconductors},\ }\href
	{https://doi.org/10.1103/PhysRevB.105.125404} {\bibfield  {journal} {\bibinfo
			{journal} {Phys. Rev. B}\ }\textbf {\bibinfo {volume} {105}},\ \bibinfo
		{pages} {125404} (\bibinfo {year} {2022})}\BibitemShut {NoStop}%
	\bibitem [{\citenamefont {Efimkin}\ \emph {et~al.}(2021)\citenamefont
		{Efimkin}, \citenamefont {Laird}, \citenamefont {Levinsen}, \citenamefont
		{Parish},\ and\ \citenamefont {Macdonald}}]{2021_Dmitry_PRB_theory}%
	\BibitemOpen
	\bibfield  {author} {\bibinfo {author} {\bibfnamefont {D.~K.}\ \bibnamefont
			{Efimkin}}, \bibinfo {author} {\bibfnamefont {E.~K.}\ \bibnamefont {Laird}},
		\bibinfo {author} {\bibfnamefont {J.}~\bibnamefont {Levinsen}}, \bibinfo
		{author} {\bibfnamefont {M.~M.}\ \bibnamefont {Parish}},\ and\ \bibinfo
		{author} {\bibfnamefont {A.~H.}\ \bibnamefont {Macdonald}},\ }\bibfield
	{title} {\bibinfo {title} {Electron-exciton interactions in the
			exciton-polaron problem},\ }\href
	{https://doi.org/10.1103/physrevb.103.075417} {\bibfield  {journal} {\bibinfo
			{journal} {Physical Review B}\ }\textbf {\bibinfo {volume} {103}},\ \bibinfo
		{pages} {075417} (\bibinfo {year} {2021})}\BibitemShut {NoStop}%
	\bibitem [{Note1()}]{Note1}%
	\BibitemOpen
	\bibinfo {note} {Our calculations follow Ref.~\cite {2017_Dmitry_PRB_theory}.
		In the recent paper~\cite {2021_Dmitry_PRB_theory}, exciton-electron
		interactions and the binding energy $\varepsilon _{\protect \mathrm {T}}$ are
		calculated microscopically and $\varepsilon _{\protect \mathrm {T}}$
		reasonably agrees with the trion binding energy evaluated within the genuine
		three-particle problem. Importantly, the doping dependence of absorption
		calculated within the two models (with microscopically derived interactions
		and ones approximated by the contact pseudopotential), are very close to each
		other.}\BibitemShut {Stop}%
	\bibitem [{\citenamefont {Zhang}\ \emph {et~al.}(2014)\citenamefont {Zhang},
		\citenamefont {Wang}, \citenamefont {Chan}, \citenamefont {Manolatou},\ and\
		\citenamefont {Rana}}]{Zhang2014}%
	\BibitemOpen
	\bibfield  {author} {\bibinfo {author} {\bibfnamefont {C.}~\bibnamefont
			{Zhang}}, \bibinfo {author} {\bibfnamefont {H.}~\bibnamefont {Wang}},
		\bibinfo {author} {\bibfnamefont {W.}~\bibnamefont {Chan}}, \bibinfo {author}
		{\bibfnamefont {C.}~\bibnamefont {Manolatou}},\ and\ \bibinfo {author}
		{\bibfnamefont {F.}~\bibnamefont {Rana}},\ }\bibfield  {title} {\bibinfo
		{title} {Absorption of light by excitons and trions in monolayers of metal
			dichalcogenide $\mathrm{Mo}{\mathrm{s}}_{2}$: Experiments and theory},\
	}\href {https://doi.org/10.1103/PhysRevB.89.205436} {\bibfield  {journal}
		{\bibinfo  {journal} {Phys. Rev. B}\ }\textbf {\bibinfo {volume} {89}},\
		\bibinfo {pages} {205436} (\bibinfo {year} {2014})}\BibitemShut {NoStop}%
	\bibitem [{\citenamefont {Adlong}\ \emph {et~al.}(2020)\citenamefont {Adlong},
		\citenamefont {Liu}, \citenamefont {Scazza}, \citenamefont {Zaccanti},
		\citenamefont {Oppong}, \citenamefont {F\"olling}, \citenamefont {Parish},\
		and\ \citenamefont {Levinsen}}]{Adlong2020}%
	\BibitemOpen
	\bibfield  {author} {\bibinfo {author} {\bibfnamefont {H.~S.}\ \bibnamefont
			{Adlong}}, \bibinfo {author} {\bibfnamefont {W.~E.}\ \bibnamefont {Liu}},
		\bibinfo {author} {\bibfnamefont {F.}~\bibnamefont {Scazza}}, \bibinfo
		{author} {\bibfnamefont {M.}~\bibnamefont {Zaccanti}}, \bibinfo {author}
		{\bibfnamefont {N.~D.}\ \bibnamefont {Oppong}}, \bibinfo {author}
		{\bibfnamefont {S.}~\bibnamefont {F\"olling}}, \bibinfo {author}
		{\bibfnamefont {M.~M.}\ \bibnamefont {Parish}},\ and\ \bibinfo {author}
		{\bibfnamefont {J.}~\bibnamefont {Levinsen}},\ }\bibfield  {title} {\bibinfo
		{title} {Quasiparticle lifetime of the repulsive fermi polaron},\ }\href
	{https://doi.org/10.1103/PhysRevLett.125.133401} {\bibfield  {journal}
		{\bibinfo  {journal} {Phys. Rev. Lett.}\ }\textbf {\bibinfo {volume} {125}},\
		\bibinfo {pages} {133401} (\bibinfo {year} {2020})}\BibitemShut {NoStop}%
	\bibitem [{\citenamefont {Darkwah~Oppong}\ \emph {et~al.}(2019)\citenamefont
		{Darkwah~Oppong}, \citenamefont {Riegger}, \citenamefont {Bettermann},
		\citenamefont {H\"ofer}, \citenamefont {Levinsen}, \citenamefont {Parish},
		\citenamefont {Bloch},\ and\ \citenamefont {F\"olling}}]{Oppong2019}%
	\BibitemOpen
	\bibfield  {author} {\bibinfo {author} {\bibfnamefont {N.}~\bibnamefont
			{Darkwah~Oppong}}, \bibinfo {author} {\bibfnamefont {L.}~\bibnamefont
			{Riegger}}, \bibinfo {author} {\bibfnamefont {O.}~\bibnamefont {Bettermann}},
		\bibinfo {author} {\bibfnamefont {M.}~\bibnamefont {H\"ofer}}, \bibinfo
		{author} {\bibfnamefont {J.}~\bibnamefont {Levinsen}}, \bibinfo {author}
		{\bibfnamefont {M.~M.}\ \bibnamefont {Parish}}, \bibinfo {author}
		{\bibfnamefont {I.}~\bibnamefont {Bloch}},\ and\ \bibinfo {author}
		{\bibfnamefont {S.}~\bibnamefont {F\"olling}},\ }\bibfield  {title} {\bibinfo
		{title} {Observation of coherent multiorbital polarons in a two-dimensional
			fermi gas},\ }\href {https://doi.org/10.1103/PhysRevLett.122.193604}
	{\bibfield  {journal} {\bibinfo  {journal} {Phys. Rev. Lett.}\ }\textbf
		{\bibinfo {volume} {122}},\ \bibinfo {pages} {193604} (\bibinfo {year}
		{2019})}\BibitemShut {NoStop}%
	\bibitem [{\citenamefont {Koschorreck}\ \emph {et~al.}(2012)\citenamefont
		{Koschorreck}, \citenamefont {Pertot}, \citenamefont {Vogt}, \citenamefont
		{Fröhlich}, \citenamefont {Feld},\ and\ \citenamefont
		{Köhl}}]{2012_M.Koschorreck_Nature_FermiPolColdAtom}%
	\BibitemOpen
	\bibfield  {author} {\bibinfo {author} {\bibfnamefont {M.}~\bibnamefont
			{Koschorreck}}, \bibinfo {author} {\bibfnamefont {D.}~\bibnamefont {Pertot}},
		\bibinfo {author} {\bibfnamefont {E.}~\bibnamefont {Vogt}}, \bibinfo {author}
		{\bibfnamefont {B.}~\bibnamefont {Fröhlich}}, \bibinfo {author}
		{\bibfnamefont {M.}~\bibnamefont {Feld}},\ and\ \bibinfo {author}
		{\bibfnamefont {M.}~\bibnamefont {Köhl}},\ }\bibfield  {title} {\bibinfo
		{title} {Attractive and repulsive fermi polarons in two dimensions},\ }\href
	{https://doi.org/10.1038/nature11151} {\bibfield  {journal} {\bibinfo
			{journal} {Nature}\ }\textbf {\bibinfo {volume} {485}},\ \bibinfo {pages}
		{619} (\bibinfo {year} {2012})}\BibitemShut {NoStop}%
	\bibitem [{\citenamefont {Scazza}\ \emph {et~al.}(2017)\citenamefont {Scazza},
		\citenamefont {Valtolina}, \citenamefont {Massignan}, \citenamefont {Recati},
		\citenamefont {Amico}, \citenamefont {Burchianti}, \citenamefont {Fort},
		\citenamefont {Inguscio}, \citenamefont {Zaccanti},\ and\ \citenamefont
		{Roati}}]{2017_Scazza_PRL_RPColdAtom}%
	\BibitemOpen
	\bibfield  {author} {\bibinfo {author} {\bibfnamefont {F.}~\bibnamefont
			{Scazza}}, \bibinfo {author} {\bibfnamefont {G.}~\bibnamefont {Valtolina}},
		\bibinfo {author} {\bibfnamefont {P.}~\bibnamefont {Massignan}}, \bibinfo
		{author} {\bibfnamefont {A.}~\bibnamefont {Recati}}, \bibinfo {author}
		{\bibfnamefont {A.}~\bibnamefont {Amico}}, \bibinfo {author} {\bibfnamefont
			{A.}~\bibnamefont {Burchianti}}, \bibinfo {author} {\bibfnamefont
			{C.}~\bibnamefont {Fort}}, \bibinfo {author} {\bibfnamefont {M.}~\bibnamefont
			{Inguscio}}, \bibinfo {author} {\bibfnamefont {M.}~\bibnamefont {Zaccanti}},\
		and\ \bibinfo {author} {\bibfnamefont {G.}~\bibnamefont {Roati}},\ }\bibfield
	{title} {\bibinfo {title} {Repulsive fermi polarons in a resonant mixture of
			ultracold li6 atoms},\ }\href
	{https://doi.org/10.1103/physrevlett.118.083602} {\bibfield  {journal}
		{\bibinfo  {journal} {Physical Review Letters}\ }\textbf {\bibinfo {volume}
			{118}},\ \bibinfo {pages} {083602} (\bibinfo {year} {2017})}\BibitemShut
	{NoStop}%
	\bibitem [{\citenamefont {Smoleński}\ \emph {et~al.}(2021)\citenamefont
		{Smoleński}, \citenamefont {Dolgirev}, \citenamefont {Kuhlenkamp},
		\citenamefont {Popert}, \citenamefont {Shimazaki}, \citenamefont {Back},
		\citenamefont {Lu}, \citenamefont {Kroner}, \citenamefont {Watanabe},
		\citenamefont {Taniguchi}, \citenamefont {Esterlis}, \citenamefont {Demler},\
		and\ \citenamefont
		{Imamoğlu}}]{2021_T.Smolenski_Nature_PolaronWignerCrystal}%
	\BibitemOpen
	\bibfield  {author} {\bibinfo {author} {\bibfnamefont {T.}~\bibnamefont
			{Smoleński}}, \bibinfo {author} {\bibfnamefont {P.~E.}\ \bibnamefont
			{Dolgirev}}, \bibinfo {author} {\bibfnamefont {C.}~\bibnamefont
			{Kuhlenkamp}}, \bibinfo {author} {\bibfnamefont {A.}~\bibnamefont {Popert}},
		\bibinfo {author} {\bibfnamefont {Y.}~\bibnamefont {Shimazaki}}, \bibinfo
		{author} {\bibfnamefont {P.}~\bibnamefont {Back}}, \bibinfo {author}
		{\bibfnamefont {X.}~\bibnamefont {Lu}}, \bibinfo {author} {\bibfnamefont
			{M.}~\bibnamefont {Kroner}}, \bibinfo {author} {\bibfnamefont
			{K.}~\bibnamefont {Watanabe}}, \bibinfo {author} {\bibfnamefont
			{T.}~\bibnamefont {Taniguchi}}, \bibinfo {author} {\bibfnamefont
			{I.}~\bibnamefont {Esterlis}}, \bibinfo {author} {\bibfnamefont
			{E.}~\bibnamefont {Demler}},\ and\ \bibinfo {author} {\bibfnamefont
			{A.}~\bibnamefont {Imamoğlu}},\ }\bibfield  {title} {\bibinfo {title}
		{Signatures of wigner crystal of electrons in a monolayer semiconductor},\
	}\href {https://doi.org/10.1038/s41586-021-03590-4} {\bibfield  {journal}
		{\bibinfo  {journal} {Nature}\ }\textbf {\bibinfo {volume} {595}},\ \bibinfo
		{pages} {53} (\bibinfo {year} {2021})}\BibitemShut {NoStop}%
\end{thebibliography}


\end{document}